# The Implementation of Hadoop-based Crawler System and Graphlite-based PageRank-Calculation In Search Engine

郭清沛 ,201428015029038, ISCAS
徐超 ,2014E8015061086, ISCAS, 宋扬 ,2014E8015061082,ISCAS

## Abstract

Nowadays, the size of the Internet is experiencing rapid growth. As of December 2014, the number of global Internet websites has more than 1 billion and all kinds of information resources are integrated together on the Internet , however,the search engine is to be a necessary tool for all users to retrieve useful information from vast amounts of web data.

Generally speaking, a complete search engine includes the crawler system, index building systems, sorting systems and retrieval system. At present there are many open source implementation of search engine, such as lucene, solr, katta, elasticsearch, solandra and so on. The crawler system and sorting system is indispensable for any kind of search engine and in order to guarantee its efficiency , the former needs to update crawled vast amounts of data and the latter requires real-time to build index on newly crawled web pages and calculae its corresponding PageRank value. It is unlikely to accomplish such huge computation tasks depending on a single hardware implementation of the crawler system and sorting system,from which aspect, the distributed cluster technology is brought to the front. In this paper, we use the hadoop Map - Reduce computing framework to implement a distributed crawler system, and use the GraphLite , a distributed synchronous graph-computing framework, to achieve the real-time computation in getting the PageRank value of the new crawled web page.

# 1.Introduction

## 1.1 Framework of Hadoop Map-Reduce

The Map - Reduce is a programming model based on Hadoop. In the Map - Reduce distributed computing framework, the programmer only writes a serial program and ensure the correctness of the serial program and then the system will complete the execution in a parallel and distributed way, which is transparent for programmers. The hadoop-based distributed computing framework is as shown in the figure 1.1.1 below:

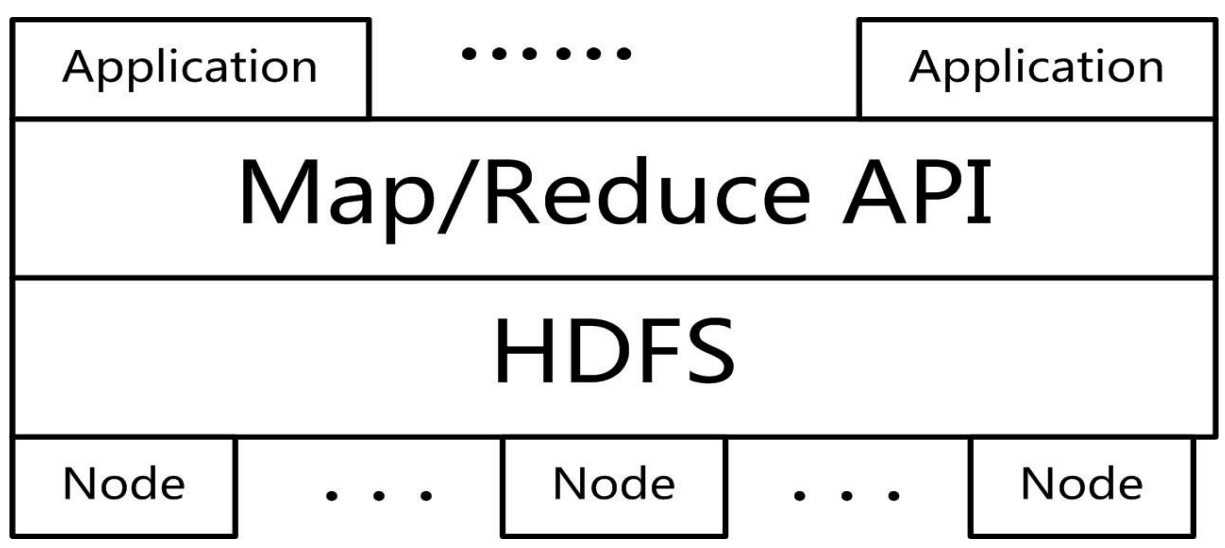

Figure 1.1.1 Hadoop Map-Reduce Framework

The lowest layer is a cluster composed of many physical nodes, each Node in the cluster is divided in logic, and the implementation of each node is just a running process so that multiple nodes can be distributed in one or more physical hosts. HDFS and MAP - REDUCE tasks run on the cluster.HDFS defines a NameNode, usually with a Secondary NameNode for redundancy backup,which are commonly responsible for storing metadata and data backup, other DataNodes are responsible for the specific file operations such as reading and writing. The Map - Reduce tasks need to run on HDFS for sharing data between different physical host nodes and storing intermediate results. When a user submitts a Map-Reduce task ,the Map - Reduce framework can decompose a task into subtasks and assign them running on corresponding nodes in cluster.In such a way distributed computation is achieved by programmers without caring about any specific distributed implementation details.

## 1.2 Framework of GraphLite

Graphlite uses a called BSP (Bulk Synchronous Processing) programming model. As shown in figure 1.2:

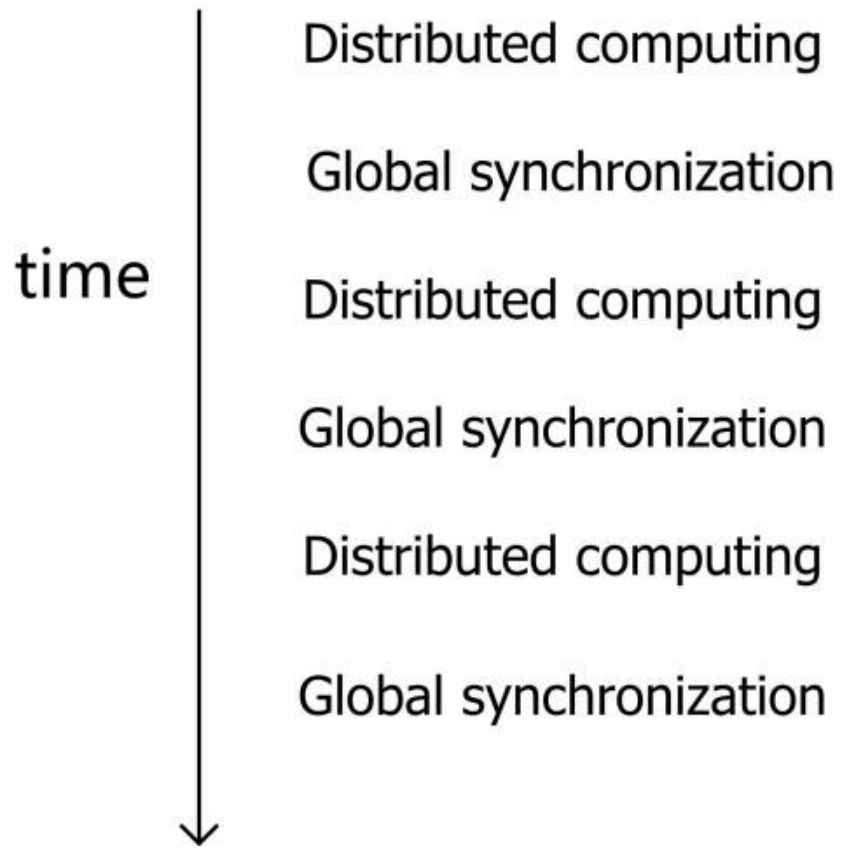

Figure 1.2.1 BSP programming model

In graphLite framework, the computation on a graph will be divide into multiple supersteps . Between two supersteps are distributed computation without any reliance , in such a way , the goal of "general serial, parallel partial" is achieved. All dependencies of nodes operation in GraphLite are classified as "data dependence" and "temporal dependence", the former can be solved through the message-sending mechanism during the initial stage of a superstep , "temporal dependence" can be resolved by a serial of sequenced SuperSteps. At the end of each SuperStep, GraphLite will collect the messages sent by all the nodes, and sending them to the corresponding destination nodes before the next superstep begins ,then starts the next round of distributed computation.

# 2.Our Implementation

## 2.1 Hadoop-based Crawler System

### 2.1.1 How we use Map-Reduce In Our Crawler-System

We use the Map-Reduce framework to implement the distributed crawler system as shown in figure 2.1.1:

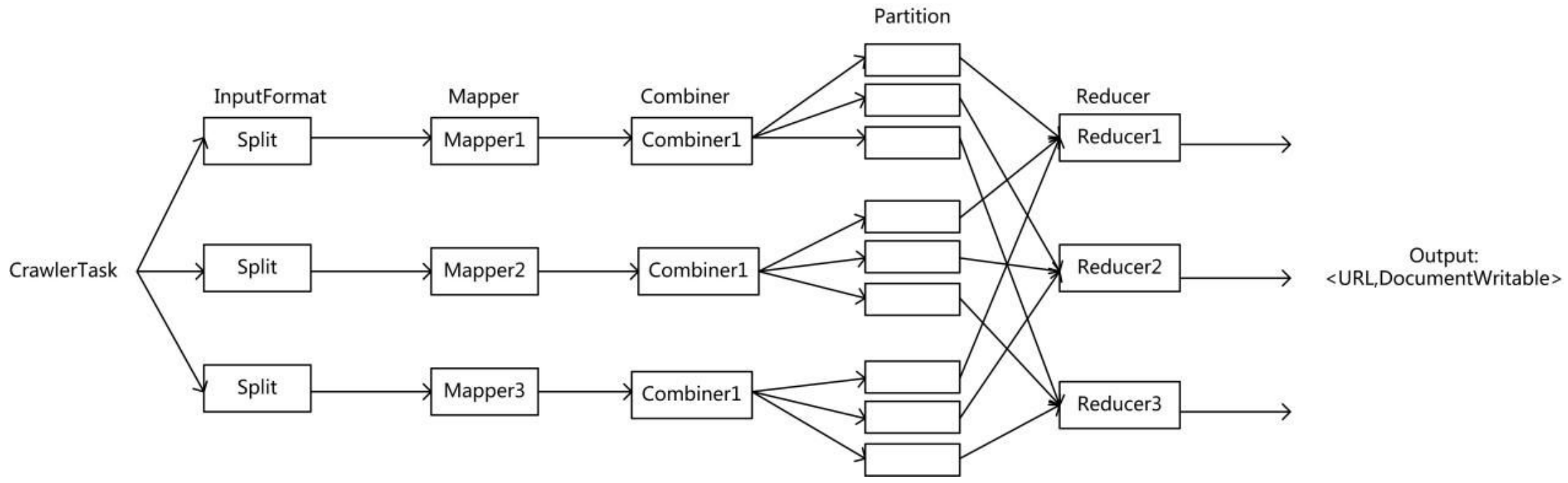

Figure2.1.1

CrawlerTask is a text file that stores the seed urls to be crawled, InputFormat is the pretreatment process before performing the of the Map operation, in which the data file will be cut into small shards, each we call it a InputSplit, defaulted with the size of 64M. Each of the InputSplits will be analyzed to a pair of <key,value>.The key of every <key,value> pair outputted by InputFormat is the starting offset of each line,while the value is the URL to be crawled.

In Map process,The Mapper class input is formatted as a set of <offset,url> pairs that analyzed by InputFormat.In our system what implemented by Map function is swapping the value of key and value in input key-value pairs. We set the url of each input-pair as the output key, whose value dedicate the extracted crawling URL .While the output value is offset. The result will be written into the intermediate files, which exists in the HDFS.

Given the thought that the overhead of communication between the nodes in Hadoop Cluster usually costs much in efficiency . We combine each Mapper output in Combiner stage of temporary files in the repeat key on the local merging, so we can reduce the amount of traffic between nodes and reduce the pressure of subsequent reducers.

What implemented in the Partitioner process is partitioning the intermediate results. According to the value of the result-key, the results could be divided into R intermediate results after the combiner process, each will be sequently processed by a Reducer. The partition algorithm we use is aimed at calculating the hash value of each URL corresponding to the host so that the URLs belonging to the same host will be partitioned into the same bin, which will be then processed by the same Reducer. So that the same host URL will be crawled exactly on the same machine. In the Reduce phase,the URLs will be used to multithreaded downloading and the crawled web pages will be written in the HDFS.

## 2.1.2 Our Running Results

The crawled results based on Map-Reduce framework is stored in the HDFS , files of which are distributed saved at different host nodes. In order to display the results conveniently, we use a database visualization tool connected HDFS to display the crawled data. Figure 2.2 shows the results crawled by our distributed Crawler-System, and Figure 2.3 roughly display the specific content of a crawled web-page.

| Id | URL | Title | keyWords | media | comment | content |
|---|---|---|---|---|---|---|
| 1 | http://sports.sina.com.cn/cba/2014-01- | 广东大胜双杀北京 易建联16+12刘晓宇立功10分3助 | 北京,广东,易建联 | 新浪体育 | 26173 | <p> 北京时间1月12日，广东主场107比90大胜北 |
| 2 | http://sports.sina.com.cn/g/seriea/2013 | 梅西逃税波及意甲!AC米兰国米被查 红黑大将+巴黎飞翼 | 意甲,梅西,国米,AC米兰,巴黎 | 新浪体育 | 359 | <p> 梅西逃税风波被曝光，他补交了1000万欧元 |
| 3 | http://sports.sina.com.cn/cba/2014-07- | 亚洲杯-遭遇罚球绝杀！国奥不敌菲律宾获得第四名 | 国奥,亚洲杯,CBA,中国男篮 | 新浪体育 | 9987 | <p> 北京时间7月19日武汉消息，2014年第五届 |
| 4 | http://2014.sina.com.cn/news/o/2014-0 | 看世界杯产生幻觉跳下二楼 建筑系博士重伤入院 | 世界杯,球迷 | 三湘都市报 | 353 | <p> 世界杯很精彩，熬夜需谨慎。千万不要为了 |
| 5 | http://sports.sina.com.cn/g/ucl/2014-10 | 曼城欧冠对手遭重罚 三场小组赛禁止球迷入场罚款20万 | 曼城,莫斯科中央陆军,欧冠, | 新浪体育 | 279 | <p class="MsoListParagraph"><span style="fon |
| 6 | http://sports.sina.com.cn/g/seriea/2014 | AC米兰购阿森纳妖锋已正式谈判 巨头密谈意大利新布冯 | AC米兰,意甲,意大利足球 | 新浪体育 | 1111 | <p> 世界杯结束之后，超级大黑马哥斯达黎加的 |
| 7 | http://sports.sina.com.cn/nba/2014-07- | 詹姆斯已现身巴西：世界杯决赛比NBA总决赛分量重 | 詹姆斯,NBA,世界杯,骑士,美 | 新浪体育 | 2801 | <p> 北京时间7月13日，据美联社报道，宣布重返 |
| 8 | http://sports.sina.com.cn/j/2014-07-31/ | 申花踢国安将留力轮换阵容？策划足协杯夺冠通亚冠 | 国安,申花,巴蒂斯塔,足协杯, | 足球 | 981 | <p> 特约记者清源、连城报道 “还不错。”7月30 |
| 9 | http://sports.sina.com.cn/j/2014-03-15/ | 央视:绿地申花困境需球迷站出来 名垦变但文化流传 | 申花,申方剑,上海 | 新浪体育 | 1637 | <!-- video_play_list_data 绿地主场1-3绿城\|1284911 |
| 10 | http://sports.sina.com.cn/g/laliga/2013 | 告诉你巴萨拉玛西亚青训有多牛 欧洲足坛的西点军校！ | | 新浪体育 | 2589 | <p> 拉玛西亚青训营不但是巴萨的青训营，也是 |
| 11 | http://sports.sina.com.cn/t/2013-03-07/ | 沃兹机智回应分手传闻 男友同证糟糕表现与恋情无关 | 沃兹尼亚奇,麦克罗伊 | 新浪体育 | 27 | <p> 北京时间3月7日消息，最近前世界第一沃兹 |
| 12 | http://sports.sina.com.cn/yayun2014/o/ | 何姿希望与李敏镐合个影：我也有点“犯花痴”(图) | 何姿,李敏镐,跳水 | 新浪体育 | 713 | <!-- HDSlide http://slide.sports.sina.com.cn/o/slic |
| 13 | http://sports.sina.com.cn/t/2014-05-28/ | 大威坦承赛前渴望与妹妹会师 称急躁心态导致失利 | 大威,女子网球 | 新浪体育 | 467 | <p> 北京时间5月28日(当地时间28日)消息，作为 |
| 14 | http://sports.sina.com.cn/j/2014-04-17/ | 人和力挺朱炯：战术改造需过程 转型过程必要付代价 | 贵州人和,朱炯,亚冠 | 足球 | 388 | <p> 特约记者王浩报道 0比1不敌川崎前锋，贵州 |
| 15 | http://sports.sina.com.cn/g/laliga/2014 | 巴萨突然官方宣布主席罗塞尔辞职 新主席任至2016 | 罗塞尔,巴萨,辞职,主席,内马 | 新浪体育 | 2983 | <!-- video_play_list_data 巴萨宣布罗塞尔辞职\|1248 |
| 16 | http://sports.sina.com.cn/g/pl/2014-01- | 4200万！切尔西后曼城也出手了 毁约金砸欧洲黑店妖星 | 曼城,拜仁,科拉罗夫,中卫 | 新浪体育 | 1515 | <p> 来自法国媒体《le10sport》的消息，曼城< |
| 17 | http://sports.sina.com.cn/cba/2014-07- | 吃止疼药也要坚持上场 男篮锋线猛将这场拼到抽筋 | 男篮,CBA,中国男篮 | 新浪体育 | 265 | <p> 北京时间7月19日消息，加时赛刚刚开始，翟 |
| 18 | http://sports.sina.com.cn/g/seriea/2013 | 尤文购特维斯协议正式达成 1200万先生已飞往体检 | 特维斯,曼城,尤文 | 新浪体育 | 5560 | <!-- video_play_list_html <table id="video_play_li |
| 19 | http://sports.sina.com.cn/cba/2014-01- | 浙江大胜佛山终结连败 斯班瑟31分朱旭航27分10板 | 佛山,浙江,朱旭航 | 新浪体育 | 76 | <p> 北京时间1月12日，2013 - 2014CBA<a cla |
| 20 | http://sports.sina.com.cn/g/ucl/2014-10 | 曼城欧冠最大罪魁是他？豪言夺冠成笑话 靠一大神保命 | 阿圭罗,曼城,欧冠,英国足球 | 足球 | 1534 | <p> <strong> 阴云遮住蓝月亮</strong></p> |
| 21 | http://2014.sina.com.cn/news/ita/2014- | 巴神：对手是乌拉圭不是苏亚雷斯 英西出局是好事 | 巴神,苏亚雷斯,出局 | 新浪体育 | 320 | <p> 北京时间今天夜间，意大利队将在世界杯小组 |
| 22 | http://sports.sina.com.cn/g/seriea/2014 | 国米甩不掉水货终于要走了 卖完悍将2000万购巴黎快马 | 巴黎,国际米兰,西尔维斯特雷 | 新浪体育 | 503 | <p> 国际米兰加快了在转会市场上的步伐，奥斯 |
| 23 | http://sports.sina.com.cn/nba/2014-07- | 骑士市值翻倍已超10亿美元 8小时内新赛季套票售罄 | 骑士,热火,詹姆斯,NBA,美职 | 新浪体育 | 19585 | <p> 北京时间7月13日，据《看台体育》报道，勒 |
| 24 | http://sports.sina.com.cn/j/2014-07-31/ | 恒大鲁能用输出球员拉拢中超盟友？ | 恒大,鲁能,亚泰,广州,中超 | 足球 | 3743 | <p> 记者陈永评述 在本土球员的交流领域，如果 |

Figure 2.2 The data list

```

<!DOCTYPE html PUBLIC "-//W3C//DTD XHTML 1.0 Transitional//EN" "http://www.w3.org/TR/xhtml1/DTD/xhtml1-transition
<!--[6,12,6977153] published at 2014-01-12 23:24:29 from #235 by 4996-->
<!-- LLTJ_MT:name ="新浪体育" -->
<!-- LLTJ_ZT:url="http://sports.sina.com.cn/cba/"; name="中国男篮&CBA职业联赛,cba";type="ZW"; -->

<html xmlns="http://www.w3.org/1999/xhtml">
<head>
<meta http-equiv="Content-type" content="text/html; charset=gb2312" />
<title>广东大胜双杀北京 易建联16+12刘晓宇立功10分3助_篮球-CBA_新浪竞技风暴_新浪网</title>
<meta name=keywords content="广东大胜双杀北京 易建联16+12刘晓宇立功10分3助,北京,广东,易建联">
<meta name=description content="广东大胜双杀北京 易建联16+12刘晓宇立功10分3助">
<meta property="og:title" content="广东大胜双杀北京 易建联16+12刘晓宇立功10分3助" />
<meta property="og:description" content="广东大胜双杀北京 易建联16+12刘晓宇立功10分3助" />
<meta property="og:url" content="http://sports.sina.com.cn/cba/2014-01-12/21246977153.shtml" />
<meta property="og:image" content="" />
<meta name="comment" content="ty:6-12-6977153">
<meta name="sudameta" content="comment_channel:ty;comment_id:6-12-6977153">
<meta name="publishid" content="427,12,6977153">
<meta name="stencil" content="PGLS000123" />
<meta name="subjectid" content="6,76,2640,2">

<meta name="sudameta" content="sinaog:1"/>
<meta http-equiv="mobile-agent" content="format=html5; url=http://dp.sina.cn/dpool/cms/jump.php?url=http%3A%2F%2F
```

Figure 2.3 The specific content of crawl pages

Distributed Crawler-System is running on the Hadoop cluster, each node in the cluster are definitely a centralized crawler, controlled by a master node to work together, so the efficiency of Distributed Crawler-System is much higher than the centralized Crawler-System. In our System ,We adopt three nodes, each node of the reducer crawling with 16 threads, 48 threads totally. Experimental results show that with the system running for 30 minutes, the size of web page stored on HDFS data is 872M (HDFS data redundancy backup number is set to 1). The single machine 16 threads run 30 minutes crawl web data of about 300M, which shows that the distributed crawler performance is much better than single node.

## 2.2 Graphlite-based PageRank-Calculation

### 2.2.1 our method to calculate PageRank

In the search engine, each web page newly crawled by Crawler-System needs to be real-time calculated its weight among all web Pages in Ranking-System,namely PageRank, according to the number of up and down links . PageRank reflects the importance of web pages, which is critical for improving the user search-experience. Here we use GraphLite , a distributed System for large-scale graph processing to calculate PageRank.

In our Crawler-System, after de-emphasis of each page, we will assign a unique Id for each of them. Each page is deemed as a vertex in graph computation System. The formula to calculate each vertex is as follows:

$$R_u = 1 - d + d\sum_{v\in B(u)} \frac{R_v}{L_v}$$

Among them:

Rv: PageRank * N of vertex N
Lv: in-degree of vertex V
B (u): out-degree of vertex u
d: web links Probability
N: the number of all pages

In the program, we initialize the weight of all the pages as 1.0, and perform iterate computation until the result comes to convergence.We enabled four worker nodes in our program, each vertex is assigned to four worker nodes based on the result of its vertex Id mod 4, the organization format of the input file is shown as Figure 2.2.1.1 .Take the input data of Worker3 for example, the input data format is shown in Figure 2.2.1.2.

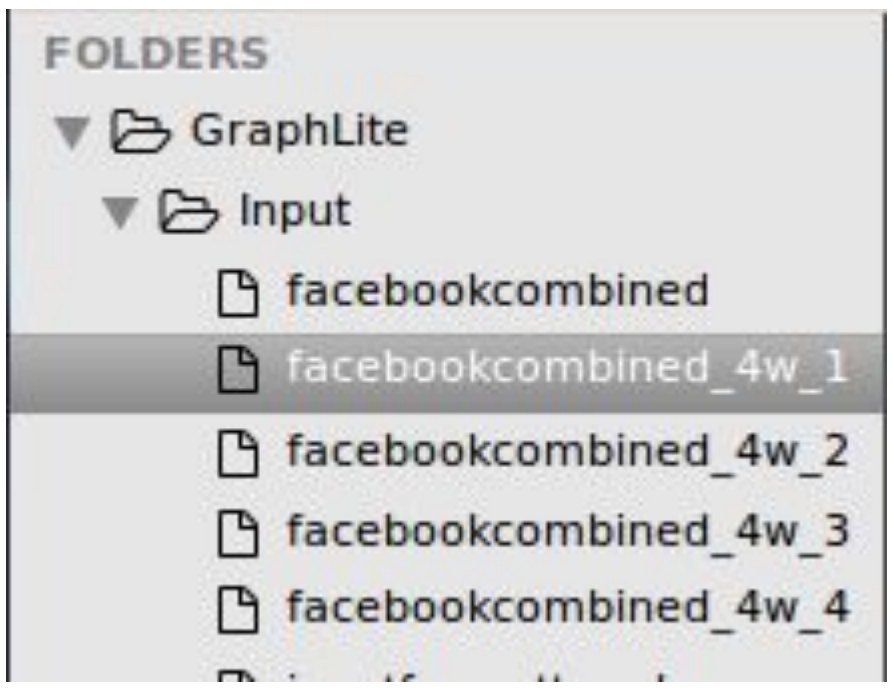

Figure 2.2.1.1

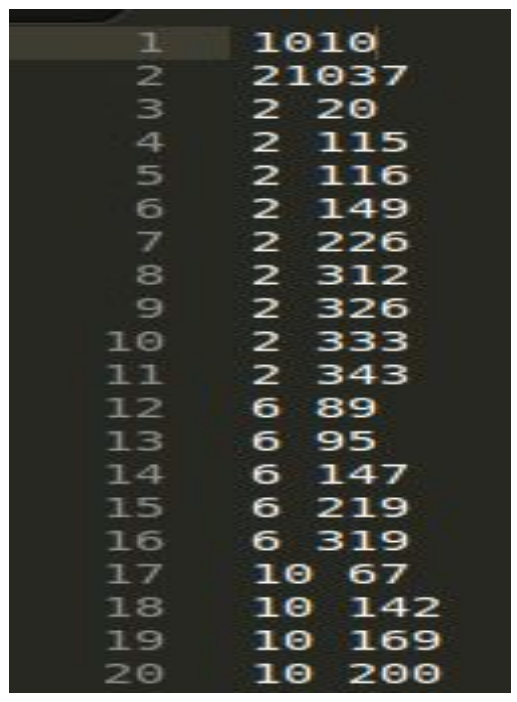

```
1010
21037
2 20
2 115
2 116
2 149
2 226
2 312
2 326
2 333
2 343
6 89
6 95
6 147
6 219
6 319
10 67
10 142
10 169
10 200
```

Figure 2.2.1.2

Figure 2.2.1.1 shows the organization format of the input file,Figure 2.2.1.2 shows the the input data format for each worker.

In Figure 2.2.1.2 ,the number 1010 in the first line represents that this worker's entry page nodes is 1010, the 21037 in the second line represents that output edge associated with entry page node amount to 21,037. Each row of data from the beginning of the third line represents an edge, for example,”2 20" represents the page, the starting id of which is 2, and the end Id is 20.

The main function of each worker is as follows:

```
class VERTEX_CLASS_NAME(): public Vertex <double, double, double>
{
public:
   void compute(MessageIterator* pmsgs) {
      double val;
      /*
      initailize  all web pages weight as 1.0
      */
      if (getSuperstep() == 0) {
        val= 1.0;
      }
      else
      {
         if (getSuperstep() >= 2)
         {
            double global_val = * (double *)getAggrGlobal(0);
            if (global_val < EPS)
            {
               /*
               if all accumulated PageRank biases are in EPS,then stop computing.
               */
               voteToHalt();
               return;
            }
         }

         double sum = 0;
         for ( ; ! pmsgs->done(); pmsgs->next() ) {
            /*
            sum up all neighbours' contributes
            */
            sum += pmsgs->getValue();
         }
         /*
         dumping factor are set as 0.85
         */
         val = 0.15 + 0.85 * sum;

         double acc = fabs(getValue() - val);
         accumulateAggr(0, &acc);
      }
      * mutableValue() = val;
      const int64_t n = getOutEdgeIterator().size();
      sendMessageToAllNeighbors(val / n);
   }
};
```

### 2.2.2 Our Running Results

During the cluster initializing state, the master node distribute all of the data to the 4 workers, as is shown in Figure 2.2.2.1.

```
hadoop@ubuntu:~/GraphLite$ start-graphlite example/PageRankVertex.so ~/GraphLite/
Input/tinygraph_4w  ~/GraphLite/Output/out
/home/hadoop/GraphLite/engine/graphlite 0 /home/hadoop/GraphLite/engine/start-wor
ker /home/hadoop/GraphLite/example/PageRankVertex.so /home/hadoop/GraphLite/Input
/tinygraph_4w /home/hadoop/GraphLite/Output/out
master run
parseCmdArg
loadUserFile
startWorkers
worker 1 cmd: ssh localhost '/home/hadoop/GraphLite/engine/start-worker /home/had
oop/GraphLite/engine/graphlite 1 /home/hadoop/GraphLite/example/PageRankVertex.so
 /home/hadoop/GraphLite/Input/tinygraph_4w /home/hadoop/GraphLite/Output/out 1'
worker 2 cmd: ssh localhost '/home/hadoop/GraphLite/engine/start-worker /home/had
oop/GraphLite/engine/graphlite 2 /home/hadoop/GraphLite/example/PageRankVertex.so
 /home/hadoop/GraphLite/Input/tinygraph_4w /home/hadoop/GraphLite/Output/out 2'
worker 3 cmd: ssh localhost '/home/hadoop/GraphLite/engine/start-worker /home/had
oop/GraphLite/engine/graphlite 3 /home/hadoop/GraphLite/example/PageRankVertex.so
 /home/hadoop/GraphLite/Input/tinygraph_4w /home/hadoop/GraphLite/Output/out 3'
worker 4 cmd: ssh localhost '/home/hadoop/GraphLite/engine/start-worker /home/had
oop/GraphLite/engine/graphlite 4 /home/hadoop/GraphLite/example/PageRankVertex.so
 /home/hadoop/GraphLite/Input/tinygraph_4w /home/hadoop/GraphLite/Output/out 4'
init
manageSuperstep
Receiver: accept all client success
received WM_BEGIN
MW_BEGIN: 1
step into sendAll
```

Figure 2.2.2.1 initialization of the cluster

Figure 2.2.2.2 shows that the program converges when proceeding to the 19th step. Computing 4039 nodes ,88,234 edges of a directed graph on Ubuntu 64bit 3.2 GHz dual-core 4 thread machine costs only 4.56s. It is almost single-node operation consuming 1 / 3! Though it seems still a little time consuming for real-time computation, but it's obvious that we can use more machines in our cluster to achieve faster computation. It is shown that the use of distributed computing that can improve the efficiency of PageRank efficiency greatly, and the use of a distributed computing architecture experiment can greatly reduce the requirements for in-memory of a single node during the computation process .

```
superstep: 19
received WM_CURSSFINISH
MW_END: 4
step into sendAll
sent MW_END to worker[1]
sent MW_END to worker[2]
sent MW_END to worker[3]
sent MW_END to worker[4]
sent MW_END
received WM_END
terminate
Receiver: closeAllSocket
Sender: closeAllSocket
elapsed: 2.574161
```

Figure 2.2.2.2 Running Results

## 3.Conclusion

In this paper, we design and implement two distributed systems to solve the real-time problem of big data processing in search engine.The Hadoop-based Crawler System and Graphlite-based PageRank-Calculation System running on a cluster are both proven highly effective than a single machine in big data processing and can be used in real Industrial production environment as a solution.